\documentclass[bibtex,twocolumn,prl,showpacs,floatfix,preprintnumbers,amsmath,amssymb,superscriptaddress,letter]{revtex4}

\usepackage{graphicx}% Include figure files
\usepackage{dcolumn}% Align table columns on decimal point

\newcommand{\numutonutau}{$\nu_{\mu}\rightarrow\nu_{\tau}$ }

\newcommand{\numu}{$\nu_{\mu}$}
\newcommand{\numubar}{$\overline{\nu}_{\mu}$}
\newcommand{\nue}{$\nu_{e}$}
\newcommand{\nuebar}{$\overline{\nu}_{e}$}

\newcommand{\cs}{$\chi^{2}$}

\newcommand{\evsq}{${\rm eV}^{2}/c^{4}$}

\newcommand{\nutau}{$\nu_{\tau}$}
\newcommand{\dmsq}{$|\Delta m^{2}_{32}|$}
\newcommand{\dmsqone}{$|\Delta m^{2}_{31}|$}

\newcommand{\sintwo}{$\rm sin^{2}(2\theta_{23})$}

\begin{document}

\preprint{FERMILAB-PUB-06-243}
\preprint{BNL-76806-2006-JA}
\preprint{hep-ex 0607088}
\title{Observation of Muon Neutrino Disappearance with the MINOS 
Detectors in the NuMI Neutrino Beam}

%\author{MINOS Authors}
%\affiliation{(The MINOS Collaboration)}

\title{Observation of muon neutrino disappearance with the MINOS detectors in the NuMI neutrino beam}

\newcommand{\Cambridge}{Cavendish Laboratory, Univ. of Cambridge, Madingley Road, Cambridge CB3 0HE, UK}
\newcommand{\FNAL}{Fermi National Accelerator Laboratory, Batavia, IL 60510}
\newcommand{\RAL}{Rutherford Appleton Laboratory, Chilton, Didcot, Oxfordshire, OX11 0QX, UK}
\newcommand{\UCL}{Dept. of Physics and Astronomy, University College London, Gower Street, London WC1E 6BT, UK}
\newcommand{\Caltech}{Lauritsen Lab, California Institute of Technology, Pasadena, CA 91125}
\newcommand{\ANL}{Argonne National Laboratory, Argonne, IL 60439}
\newcommand{\Athens}{Department of Physics, University of Athens, GR-15771 Athens, Greece}
\newcommand{\NTUAthens}{Dept. of Physics, National Tech. Univ. of Athens, GR-15780 Athens, Greece}
\newcommand{\Benedictine}{Physics Dept., Benedictine University, Lisle, IL 60532}
\newcommand{\BIMC}{Dept. of Rad. Oncology, Beth Israel Med. Center, New York, NY 10003}
\newcommand{\BNL}{Brookhaven National Laboratory, Upton, NY 11973}
\newcommand{\Beijing}{Inst. of High Energy Physics, Chinese Academy of Sciences, Beijing 100039, China}
\newcommand{\CdF}{APC -- Coll\`{e}ge de France, 11 Place Marcelin Berthelot, F-75231 Paris Cedex 05, France}
\newcommand{\Cleveland}{Cleveland Clinic, Cleveland, OH 44195}
\newcommand{\Columbia}{Physics Department, Columbia University, New York, NY 10027}
\newcommand{\GEHealth}{GE Healthcare, Florence SC 29501}
\newcommand{\DOE}{Div. of High Energy Physics, U.S. Dept. of Energy, Germantown, MD 20874}
\newcommand{\Harvard}{Dept. of Physics, Harvard University, Cambridge, MA 02138}
\newcommand{\HolyCross}{Holy Cross College, Notre Dame, IN 46556}
\newcommand{\IIT}{Physics Division, Illinois Institute of Technology, Chicago, IL 60616}
\newcommand{\Indiana}{Physics Department, Indiana University, Bloomington, IN 47405}
\newcommand{\ITEP}{High Energy Exp. Physics Dept., Inst. of Theor. and Exp. Physics, 
  B. Cheremushkinskaya, 25, 117218 Moscow, Russia}
\newcommand{\JMU}{Physics Dept., James Madison University, Harrisonburg, VA 22807}
\newcommand{\JINR}{Joint Inst. for Nucl. Research, Dubna, Moscow Region, RU-141980, Russia}
\newcommand{\LASL}{Nucl. Nonprolif. Div., Threat Reduc. Dir., Los Alamos National Laboratory, Los Alamos, NM 87545}
\newcommand{\LBL}{Physics Div., Lawrence Berkeley National Laboratory, Berkeley, CA 94720}
\newcommand{\Lebedev}{Nuclear Physics Dept., Lebedev Physical Inst., Leninsky Prospect 53, 117924 Moscow, Russia}
\newcommand{\LLL}{Lawrence Livermore National Laboratory, Livermore, CA 94550}
\newcommand{\MIT}{Lincoln Laboratory, Massachusetts Institute of Technology, Lexington, MA 02420}
\newcommand{\Minnesota}{School of Physics and Astronomy, University of Minnesota, Minneapolis, MN 55455}
\newcommand{\Crookston}{Math, Science and Technology Dept., Univ. of Minnesota -- Crookston, Crookston, MN 56716}
\newcommand{\Duluth}{Dept. of Physics, Univ. of Minnesota -- Duluth, Duluth, MN 55812}
\newcommand{\Oxford}{Sub-dept. of Particle Physics, Univ. of Oxford,  Denys Wilkinson Bldg, Keble Road, Oxford OX1 3RH, UK}
\newcommand{\PSU}{Dept. of Physics, Pennsylvania State Univ., University Park, PA 16802}
\newcommand{\Pittsburgh}{Dept. of Physics and Astronomy, Univ. of Pittsburgh, Pittsburgh, PA 15260}
\newcommand{\IHEP}{Inst. for High Energy Physics, Protvino, Moscow Region RU-140284, Russia}
\newcommand{\RoyalH}{Physics Dept., Royal Holloway, Univ. of London, Egham, Surrey, TW20 0EX, UK}
\newcommand{\Carolina}{Dept. of Physics and Astronomy, Univ. of South Carolina, Columbia, SC 29208}
\newcommand{\SLAC}{Stanford Linear Accelerator Center, Stanford, CA 94309}
\newcommand{\Stanford}{Department of Physics, Stanford University, Stanford, CA 94305}
\newcommand{\Sussex}{Dept. of Physics and Astronomy, University of Sussex, Falmer, Brighton BN1 9QH, UK}
\newcommand{\TexasAM}{Physics Dept., Texas A\&M Univ., College Station, TX 77843}
\newcommand{\Texas}{Dept. of Physics, Univ. of Texas, 1 University Station, Austin, TX 78712}
\newcommand{\TechX}{Tech-X Corp, Boulder, CO 80303}
\newcommand{\Tufts}{Physics Dept., Tufts University, Medford, MA 02155}
\newcommand{\UNICAMP}{Univ. Estadual de Campinas, IF-UNICAMP, CP 6165, 13083-970, Campinas, SP, Brazil}
\newcommand{\USP}{Inst. de F\'{i}sica, Univ. de S\~{a}o Paulo,  CP 66318, 05315-970, S\~{a}o Paulo, SP, Brazil}
\newcommand{\Washington}{Physics Dept., Western Washington Univ., Bellingham, WA 98225}
\newcommand{\WandM}{Dept. of Physics, College of William \& Mary, Williamsburg, VA 23187}
\newcommand{\Wisconsin}{Physics Dept., Univ. of Wisconsin, Madison, WI 53706}
\newcommand{\deceased}{Deceased.}

\affiliation{\ANL}
\affiliation{\Athens}
\affiliation{\Benedictine}
\affiliation{\BNL}
\affiliation{\Caltech}
\affiliation{\Cambridge}
\affiliation{\UNICAMP}
\affiliation{\Beijing}
\affiliation{\CdF}
\affiliation{\Columbia}
\affiliation{\FNAL}
\affiliation{\Harvard}
\affiliation{\IIT}
\affiliation{\Indiana}
\affiliation{\IHEP}
\affiliation{\ITEP}
\affiliation{\JMU}
\affiliation{\JINR}
\affiliation{\Lebedev}
\affiliation{\LLL}
\affiliation{\UCL}
\affiliation{\Minnesota}
\affiliation{\Duluth}
\affiliation{\Oxford}
\affiliation{\Pittsburgh}
\affiliation{\RAL}
\affiliation{\USP}
\affiliation{\Carolina}
\affiliation{\Stanford}
\affiliation{\Sussex}
\affiliation{\TexasAM}
\affiliation{\Texas}
\affiliation{\Tufts}
\affiliation{\Washington}
\affiliation{\WandM}
\affiliation{\Wisconsin}

\author{D.~G.~Michael}
\altaffiliation{\deceased}
\affiliation{\Caltech}

\author{P.~Adamson}
\affiliation{\FNAL}
\affiliation{\UCL}
\affiliation{\Sussex}

\author{T.~Alexopoulos}
%\altaffiliation[Now at\ ]{\NTUAthens .}
\affiliation{\Wisconsin}

\author{W.~W.~M.~Allison}
\affiliation{\Oxford}

\author{G.~J.~Alner}
\affiliation{\RAL}

\author{K.~Anderson}
\affiliation{\FNAL}

\author{C.~Andreopoulos}
\affiliation{\RAL}
\affiliation{\Athens}

\author{M.~Andrews}
\affiliation{\FNAL}

\author{R.~Andrews}
\affiliation{\FNAL}

\author{K.~E.~Arms}
\affiliation{\Minnesota}

% ...

\collaboration{The MINOS Collaboration}
\noaffiliation

\date{\today}

\begin{abstract}
This letter reports results from the MINOS experiment based on
      its initial exposure to neutrinos from the Fermilab NuMI beam.  The
      rates and energy spectra of charged current \numu{} interactions
      are compared in two detectors located along the beam axis at distances
      of 1\,km and 735\,km.  
With $1.27\times 10^{20}$ 120\,GeV protons incident
      on the NuMI target, 215 events with energies below
      30\,GeV are observed at the Far Detector, compared to an
      expectation of $336\pm 14.4$ events.
The data
are consistent with \numu{} disappearance via oscillations with
\dmsq~=2.74$^{+0.44}_{-0.26}\times 10^{-3}$\,\evsq{} and 
\sintwo~$>$ 0.87 (68\% C.L.).
\end{abstract}
\maketitle
There is now substantial evidence\, 
\cite{ref:osc1,ref:osc2,ref:osc3,ref:osc4,ref:osc5,ref:osc6,ref:osc7,ref:osc8} 
that the proper description of neutrinos involves
a rotation between mass and flavor eigenstates governed by
the 3$\times$3 PMNS 
matrix\,\cite{ref:pmns1,ref:pmns2}.
The parameters of this 
mixing matrix, three angles and a phase, as well as the mass differences
between the three mass eigenstates must be determined experimentally.
The Main Injector Neutrino Oscillation Search (MINOS) 
experiment has been designed to study the flavor composition of a 
beam of muon neutrinos as it travels between
the Near Detector (ND) at Fermi National Accelerator Laboratory 
at 1\,km from the target, 
and the Far Detector (FD) in the Soudan iron mine in Minnesota at 
735\,km from the target. 
From the comparison of the reconstructed neutrino energy spectra
at the near and far locations the oscillation parameters
\dmsq{}
and \sintwo{} are extracted.

The Neutrinos at the Main Injector (NuMI) neutrino beam is produced using
120\,GeV protons from the Main Injector. The protons are delivered
in 10\,$\mu$s spills with up to 3.0$\times 10^{13}$ protons per spill.
The extracted protons are bent downward by 
3.3$^{\circ}$ to point at the MINOS 
detectors. The global positioning system (GPS)
defined the survey beam direction to 
within 12\,m of the FD\,\cite{ref:pointing}.
Positively charged particles produced by the proton beam
in the 95.4\,cm long target
(mainly $\pi^{+}$ and $K^{+}$) are focused by two pulsed parabolic 
horns spaced 10\,m apart
and are then allowed to decay in a 675\,m long, 2\,m diameter, 
evacuated decay pipe\,\cite{ref:target}. The proton beam\,\cite{ref:pribeam}
and tertiary muon beam\,\cite{ref:monitoring} are
monitored on a pulse-by-pulse 
basis.
The target position relative to the first horn 
and the horn current are variable\,\cite{ref:horns}. 
For the majority of the running 
period described here, the  
target was inserted 50.4\,cm into the first 
horn to maximize neutrino
production in the 1-3\,GeV energy range. 
A total of 1.27$\times 10^{20}$ protons on target (POT) were taken
in this position and used for the oscillation analysis.
The charged current (CC) neutrino event yields at the ND are predicted to be 
92.9\% \numu, 5.8\% \numubar, 
1.2\% \nue{} and 0.1\% \nuebar. 
The data described here were recorded between May 2005 and February 2006. 
The average livetime of the FD was 
99.0\% during this period.
About one third of the total ND events provided a sufficiently large 
dataset for this analysis of $\sim$10$^{6}$ events which were 
sampled throughout the run period.

Both MINOS detectors \,\cite{ref:minos1} are steel-scintillator 
tracking calorimeters \,\cite{ref:minos2} with 
toroidal magnetic fields averaging 1.3\,T\,\cite{ref:steel}. The steel
plates are 2.54\,cm thick. 
The scintillator planes are comprised of 4.1\,cm wide 
and 1\,cm thick plastic strips. 
Each plane is oriented at 
45$^{\circ}$ from vertical and at 90$^{\circ}$ with respect to
its neighbors. 
The light from the scintillator strips is transported to the
multi-anode photomultiplier tubes (PMT) by embedded 1.2\,mm diameter 
wavelength shifting (WLS) fibers. 
In order to cancel the majority of the
uncertainties in the modeling of neutrino interactions and detector
response, the two MINOS detectors are as similar as possible. 
For example, both detectors yield $6-7$ photoelectrons (PEs) per plane for
normally incident minimum ionizing particles.
However, the
data rate in the ND is $\sim10^{5}$ times larger than in the FD which
has dictated certain design differences between them.

The 5.4\,kton FD, 705\,m underground,
has 484 octagonal, 8\,m wide 
instrumented planes read out at both ends via Hamamatsu M16 
PMTs\,\cite{ref:m16}. 
Eight WLS fibers from strips in the same plane, 
separated from each other by about 1\,m,  
are coupled to each pixel. 
The coupling pattern is 
different at the two ends to allow resolution of ambiguities. 

The 0.98\,kton ND, 103\,m underground, 
has 282 irregular
4$\times$6\,m$^{2}$ octagonal planes. Its geometry optimizes the containment
of hadronic showers and provides sufficient flux return to
achieve a magnetic field similar to the FD. 
Each strip is coupled via a WLS fiber to one pixel of a Hamamatsu 
M64 PMT\,\cite{ref:m64}. 
The ND readout continuously integrates the PMT charges with a sampling rate
of 53.1\,MHz to allow discrimination between successive Main Injector
RF buckets.

The data aquisition\,\cite{ref:daq1,ref:daq2,ref:daq3}
accepts data above a threshold of 0.25\,PEs.
In the FD, the online trigger conditions
require a hit within 100\,$\mu$s
centered on the time of the expected beam spill,
at least 20\,PEs inside a four plane window, or 4 hits
in 5 consecutive planes.
In the ND, all the data taken during the beam spill are retained. 
The trigger efficiency for both
detectors exceeds \,99.5\% for neutrino events with
visible energy above 0.5\,GeV.

The detectors are calibrated using an {\em in-situ} 
light injection system\,\cite{ref:li} and cosmic ray muons. 
LED generated light signals
are distributed to all the WLS fibers to
track gain changes in the PMTs and electronics. 
The energy deposited by through-going muons is used to 
equalize the response of all the scintillator strips.
After calibration, remaining time and position dependent variations in the 
responses of the detectors result in an uncertainty in the relative 
energy scale between the two detectors of 2\%.
The overall energy scale for single hadrons and electrons was determined
from the results of a test-beam experiment 
using a small, unmagnetized copy of the MINOS
calorimeters (CalDet)\,\cite{ref:caldet}. 
Stopping muons are then used to relate the results from CalDet
to the response of the ND and FD. From these studies, the uncertainty
on the absolute hadronic energy scale is estimated to be 
6\%.

The simulation of the production and detection of neutrinos
commences 
with a model of hadron production in the target using
FLUKA05\,\cite{ref:fluka}, which has 
uncertainties at the 20-30\% level stemming from a lack
of relevant thick target hadron production data.
The shower products are
transported through the horn focusing system and decayed in 
a GEANT3\,\cite{ref:geant} simulation that includes the horns, 
beamline material
and the decay pipe. 
The neutrino event generator, NEUGEN3\,\cite{ref:nugen}, is tuned to existing 
CC cross-section data where present uncertainties below 10\,GeV are at the
20\% level. 
The products of the neutrino interaction are propagated out of
the iron nucleus using the 
INTRANUKE\,\cite{ref:intranuke} code from 
within NEUGEN3. Some of the energy of absorbed pions is transferred to 
clusters of nucleons as motivated by Ref.\,\cite{ref:ransome}.
The response of the detector is simulated using GEANT3 with 
the GCALOR\,\cite{ref:gcalor} model of hadronic interactions.
The final step in the simulation chain involves photon generation, 
propagation and transmission through the WLS fiber and conversion 
to photoelectrons in the PMTs.

In CalDet, GEANT3 with GCALOR is found to reproduce the hadronic 
and electromagnetic (EM) 
responses of the detector to single particles to 4\% and 2\%, respectively.
Below 10\,GeV, the hadronic energy
resolution was measured to be 
56\%/$\sqrt{E\rm{[GeV]}}\oplus 2\%$\,\cite{ref:mike}
and the EM resolution was measured to be
21.4\%/$\sqrt{E\rm{[GeV]}}\oplus 4.1\%/E\rm{[GeV]}$\,\cite{ref:trish}.  
The muon energy resolution 
$\Delta E_{\mu}/E_{\mu}$ 
varies smoothly from 6\% for E$_\mu$ above 1 GeV 
where most tracks are contained and measured by range, to
13\% at high energies, where the curvature measurement is primarily used.

The initial step in the reconstruction of the FD data is the removal of the 
eightfold hit-to-strip ambiguity using information from both strip ends. 
In the ND, timing and spatial information is first used to separate individual
neutrino interactions from the same spill. Subsequently, tracks are 
found and fitted, and showers are reconstructed, in the same way in both 
detectors. 
For \numu{} CC events, the total reconstructed event 
energy is obtained
by summing the muon energy and the visible energy of the hadronic system.

The FD data set was left blind until the selection
procedure had been defined and the prediction of the unoscillated spectrum
was understood. 
The blinding procedure hid a substantial fraction of the FD 
events with the precise fraction and energy spectrum of the 
hidden sample unknown.
Events are pre-selected in both detectors,
by requiring total reconstructed energy below 30\,GeV and a 
negatively charged track to suppress events that
originate from 
$\pi^{-}$ or $K^{\pm}$.
The track vertex must be
within a fiducial volume such that cosmic rays 
are rejected
and the hadronic energy of the event is contained within the 
volume of the detector. 
The event time must fall within a 50\,$\mu$s
window around the spill time.
Cosmic ray background is suppressed further in the FD 
by requiring the track to point
within 53$^{\circ}$ of the neutrino beam direction. 
The pre-selected \numu{} event sample is predominantly CC with a 8.6\% 
NC background estimated from Monte Carlo (MC) simulations.
The fiducial mass of the FD (ND) is 72.9\% (4.5\%) of the total
detector mass.

\begin{figure}
\includegraphics[width=3.25in]{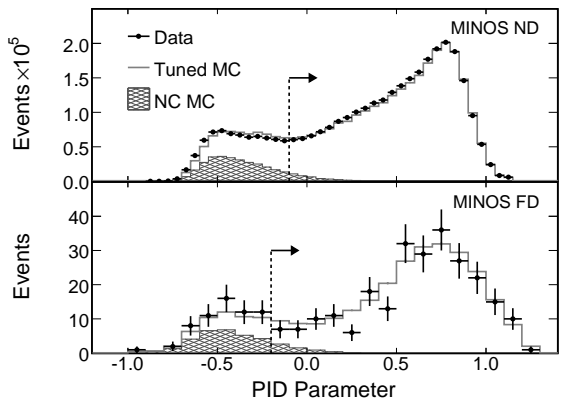}
\caption{\label{fig:pidn} 
Data and tuned MC predictions for the PID variable in the ND (top) and 
FD (bottom). The arrows depict the positions of the ND and FD selection cuts.
The FD MC distribution for CC
events uses the best fit parameters discussed in the text.}
\end{figure}

A particle identification parameter (PID) incorporating 
probability density functions for the event length, the 
fraction of energy contained 
in the track and the average track pulse height per plane provides
separation of \numu{} CC and NC events.
The PID
is shown in Fig.\,\ref{fig:pidn}  for ND and 
FD data overlaid with simulations of NC and CC events after the beam
reweighting procedure described below. 
Events with PID above -0.2 (FD) and -0.1 (ND) are selected
as being predominantly CC in origin.
These values were optimized for both detectors such that the resulting
purity of each sample is about 98\%. 
The efficiencies for selecting \numu{} CC events in the fiducial
volume with energy below 30\,GeV are 74\% (FD) and 67\% (ND).
From the absence of any events less than 20\,$\mu$s 
before and less than 30\,$\mu$s
after the spill time, the remaining non-beam related background 
in the FD is estimated to be less than 0.5 events (68\% C.L.). 
Background from \numu{} interactions
in the rock surrounding the FD is estimated from MC to be below 0.4 
(68\% C.L.) events.
The corresponding backgrounds in the ND are negligible.

\begin{figure}
\includegraphics[width=3.35in]{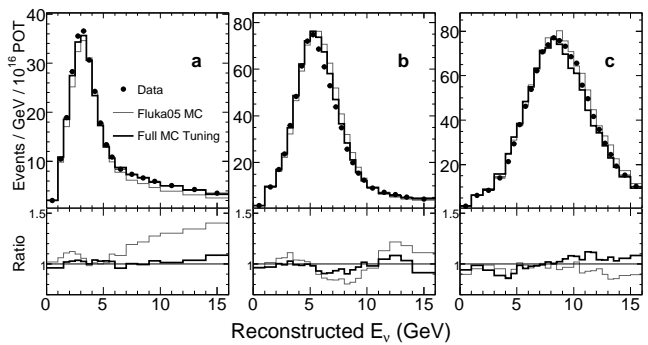}
\caption{\label{fig:tuning} 
Energy spectra in the MINOS ND for three of the six beam configurations before
and after the 15 parameter beam tuning procedure. 
The target location was modified to produce
the different spectra: a) nominal, b) target at 90\,cm from nominal, c)
target at 240\,cm from nominal. 
The lower inset shows the ratio of data to MC before and after tuning.
}
\end{figure}

To constrain hadron production,  
a series of six runs of similar exposure was taken where the position 
of the target and the magnitude of the horn magnetic field were varied.  
Comparisons of the ND energy spectra with MC simulations, shown in 
Fig.\,\ref{fig:tuning}, 
showed an energy dependent discrepancy that changed with the beam 
settings.  
This implied
beam modeling, rather than detector or cross-section
effects, was the primary cause. 
To bring the MC into better agreement with the data, a tuning 
of the beam MC was performed 
in which pion production off the target was smoothly varied in
transverse and longitudinal momentum 
with respect to the FLUKA05 input, as was the overall kaon yield.  In 
addition, the potential systematic effects of the beam focusing, 
NC background, \numu{} energy scale and offset  
were allowed to vary. All of these parameters were found to lie 
within two standard deviations of their nominal values.
Fig.\,\ref{fig:tuning} shows the 
effect of the full beam parameter tuning for the spectra corresponding to 
three different target positions.  
The resulting agreement is improved in all beams
across the 1-30\,GeV neutrino energy region. 

The measurement of the energy spectrum at the ND is used
to predict the unoscillated spectrum
at the FD.
The oscillation hypotheses are then tested relative to this 
prediction. 
The prediction must take into account the ND and FD 
spectral differences
that are present, even in the absence of oscillations, due to 
pion decay kinematics and beamline geometry.
These introduce a ND/FD shape difference of up to $\sim$20\% 
on either side of the peak.

There are two distinct approaches to the beam extrapolation. 
The {\em ND Fit} method focuses on minimizing the
remaining ND data and MC differences by modifying MC parameters
associated with neutrino interactions and detector response.
The FD MC is then re-weighted with the best-fit values of these parameters.
For the results presented in this paper, the
{\em Beam Matrix} method\,\cite{ref:adam} is used, 
in which agreement between MC and data is much less 
important because the ND data are used to 
measure all the effects common to both detectors, such as beam modeling,
neutrino interactions and detector response. It 
utilizes the beam simulation to derive a transfer matrix that 
relates \numu{}\,s in the two detectors via their parent hadrons.
The matrix element
$M_{ij}$ gives the relative probability that the distribution of 
secondary hadrons which
produce \numu\,s of energy $E_{i}$ in the ND will give \numu\,s of energy
$E_{j}$ in the FD. 
The ND reconstructed event energy spectrum is translated into a 
flux by first correcting
for the simulated ND acceptance and then dividing by the
calculated cross-sections for each energy bin.
This flux is multiplied by the matrix to yield the predicted,
unoscillated FD flux. After the inverse correction for cross-section and
FD acceptance, the predicted FD visible energy spectrum is obtained. 

\begin{figure}
\includegraphics[width=3.25in]{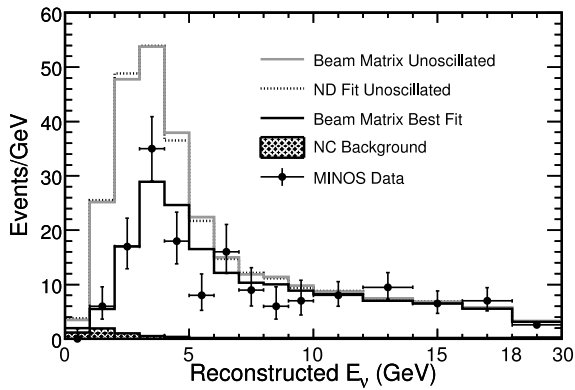}
\caption{\label{fig:spectra} 
Comparison of the Far Detector spectrum with predictions 
for no oscillations for both analysis methods and 
for oscillations with the best-fit parameters from the 
Beam Matrix extrapolation
method. The estimated NC background is also shown. The last energy bin
contains events between 18-30\,GeV.
}\end{figure}

In total, 215 events are observed below 30\,GeV compared to the
unoscillated expectation of 336.0$\pm$14.4.
The error is due to the systematic uncertainties described below.
In the region below 10\,GeV, 122 events are observed compared to the
expectation of 238.7$\pm$10.7.
The observed energy spectrum is shown along with the 
predicted spectra for both extrapolation methods in Fig.\,\ref{fig:spectra}.

Under the assumption that the observed deficit is due 
to \numutonutau{} oscillations\,\cite{ref:chooz,ref:pverde,ref:k2k},
a fit is performed to the parameters \dmsq{} and \sintwo{}
using the expression for the \numu{} survival probability:
\begin{equation}
  P(\nu_{\mu} \rightarrow \nu_{\mu})=1-\rm{sin}^2(2\theta_{23}) \rm{sin}^2(1.27\Delta {\it m}^2_{32}\frac{\it L}{\it E})
\label{eq:osc}
\end{equation}
where $L$[km] is the distance from the target, 
$E$[GeV] is the neutrino energy, and 
\dmsq{}\,\cite{ref:fogli}
is measured in \evsq.
The FD data are binned in reconstructed event
energy and the observed number of events 
in each bin is compared to the expected number of events for this oscillation
hypothesis. The best fit parameters are those
which minimize \cs=$-2\ln\lambda$ where $\lambda$ is the likelihood ratio:
\begin{equation}
\chi^{2}=\sum_{nbins}\left(2(e_{i}-o_{i})+2 o_{i}\ln(o_{i}/e_{i})\right) + \sum_{nsys}\frac{\Delta s_{j}^{2}}{\sigma_{s_{j}}^{2}}
\label{eq:ll}
\end{equation}
where $o_i$ and $e_i$ are 
the observed and expected numbers
of events in bin $i$, and the $\Delta s_j^2/\sigma_{s_j}^2$ are the penalty
terms for nuisance parameters associated with the systematic uncertainties.
The expected number of events depends on \dmsq{}, \sintwo{} and the $s_j$.
The choice of these systematic effects and 
their estimated uncertainties
are described below. 
The $e_{i}$ include 
the small contribution
from selected \nutau{} events produced in the oscillation process.

The effects of different systematic uncertainties were evaluated
by modifying the MC and performing a fit on this in place of the data.
The differences between the fitted values obtained
with the modified and unmodified MC are listed in Table\,\ref{tab:systematics}.
The largest effects are:
(a) The uncertainty in the fiducial mass in both detectors,
uncertainty in the event selection efficiency 
and the POT counting accuracy gives a 4\% uncertainty on the 
predicted FD event rate.
(b) The absolute hadronic energy scale from a combination of test beam 
measurements and calibration accuracy is known to 6\% as discussed above. 
This is added in quadrature to the
uncertainty in the effect of intra-nuclear re-scattering estimated at 
$\pm10\%$ of the hadronic energy. The total hadronic energy scale uncertainty
is therefore $\pm11\%$.
(c) The NC component was varied in a fit to the PID data distribution 
in six energy bins in the ND. A 50\% uncertainty was estimated
to encompass the differences between the fit and NC MC.
At the current level  of statistics, uncertainties from
CC cross-sections, muon momentum, relative ND/FD energy calibration,
remaining beam uncertainties and reconstruction were found
to be negligible. As an example, in the absence of 
any beam tuning, the best fit value only shifts by 0.2$\times 10^{-5}$\evsq.

\begin{figure}
\includegraphics[width=2.85in]{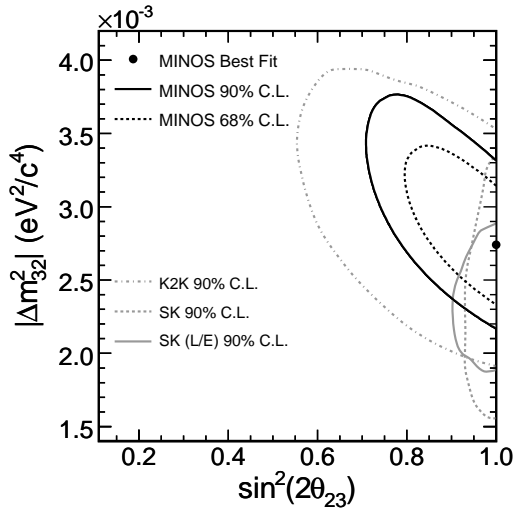}
\caption{\label{fig:contour} 
Confidence intervals for the fit using the Beam Matrix method 
including systematic errors.
Also shown are the contours from the previous highest 
precision experiments\,\cite{ref:osc1,ref:osc2,ref:osc5}.}
\end{figure}

In fitting the data to Eqn.\,\ref{eq:osc}, \sintwo{} was constrained to
lie in the physical region and the main systematic uncertainties 
((a), (b) and (c) in Table\,\ref{tab:systematics})
were included as the nuisance parameters. 
The resulting 68\% and 90\% confidence intervals are shown
in Fig.\,\ref{fig:contour} as determined from 
$\Delta$\cs=2.3 and 4.6, respectively\,\cite{ref:FC}.
The best fit value for \dmsq{} is
\dmsq\,=(2.74\,$^{+0.44}_{-0.26})\times 10^{-3}$\,\evsq{}
and \sintwo\,$>$\,0.87 at 68\% C.L.\,\cite{ref:dof} 
with a fit probability of 8.9\%.
At 90\% C.L. 
(2.31$<$\dmsq$<3.43)\times 10^{-3}$\,\evsq{}, 
and \sintwo\,$>$\,0.78.
The data and best fit MC are shown in Fig.\,\ref{fig:spectra}.
At the best fit value, the MC predicts 0.76 \nutau{} 
events in the final sample.
If the fit is not constrained to be within the physical region, 
\dmsq{}=2.72\,$\times 10^{-3}$\,\evsq{} and \sintwo{}= 1.01,
with a 0.2 decrease in \cs{}.
With additional data, it is expected that the systematic 
uncertainties will be reduced.
\begin{table}
\begin{ruledtabular}
\begin{tabular}{lcc}
Uncertainty & \dmsq& \sintwo \\
&(10$^{-3}$ eV$^2/c^{4}$)&\\ 
\hline
(a)\,Normalization ($\pm$  4\%)		& 0.05 & 0.005 \\
(b)\,Abs. hadronic E scale ($\pm$ 11\%) & 0.06 & 0.048 \\
(c)\,NC contamination ($\pm$ 50\%)  	& 0.09 & 0.050 \\
All other systematics 		        & 0.04 & 0.011 \\
\end{tabular}
\end{ruledtabular}
\caption{\label{tab:systematics}
Sources of systematic uncertainties in the
measurement of \dmsq{} and \sintwo{}.
The values of \dmsq{} and \sintwo{} 
used in the systematic MC study were the best fit values from the data. 
The values are the average shifts for varying the parameters in both
directions without imposing constraints on the fit. 
Correlations between the systematic 
effects are not taken into account.
}
\end{table}

This work was supported by the US DOE; the UK PPARC; the US NSF;
the State and University of Minnesota; 
the University of Athens, Greece and 
Brazil's FAPESP and CNPq.
We are grateful to the Minnesota 
Department of Natural Resources,
the crew of the Soudan Underground  
Laboratory, and the staff of Fermilab for their contribution to this effort.

\end{document}